# AI Techniques for Cone Beam Computed Tomography in Dentistry: Trends and Practices

Saba Sarwar
*Department of Computer Science*
*Jamia Millia Islamia*
New Delhi, India
saba188684@st.jmi.ac.in

Suraiya Jabin
*Department of Computer Science*
*Jamia Millia Islamia*
New Delhi, India
sjabin@jmi.ac.in

*Abstract*— **Cone-beam computed tomography (CBCT) is a popular imaging modality in dentistry for diagnosing and planning treatment for a variety of oral diseases with the ability to produce detailed, three-dimensional images of the teeth, jawbones, and surrounding structures. CBCT imaging has emerged as an essential diagnostic tool in dentistry. CBCT imaging has seen significant improvements in terms of its diagnostic value, as well as its accuracy and efficiency, with the most recent development of artificial intelligence (AI) techniques. This paper reviews recent AI trends and practices in dental CBCT imaging. AI has been used for lesion detection, malocclusion classification, measurement of buccal bone thickness, and classification and segmentation of teeth, alveolar bones, mandibles, landmarks, contours, and pharyngeal airways using CBCT images. Mainly machine learning algorithms, deep learning algorithms, and super-resolution techniques are used for these tasks. This review focuses on the potential of AI techniques to transform CBCT imaging in dentistry, which would improve both diagnosis and treatment planning. Finally, we discuss the challenges and limitations of artificial intelligence in dentistry and CBCT imaging.**

## I. INTRODUCTION

Artificial intelligence (AI) has become more prevalent in dentistry over the past few years, with the potential to revolutionize the field. Using image analysis, natural language processing, and decision-making as examples, machine learning enables machines to implement tasks that would normally demand intelligence from humans. Artificial intelligence (AI), also known as machine intelligence, is the ability of computers to carry out operations that typically require human intelligence, such as image analysis, natural language processing, and decision-making.

Machine Learning (ML) is a subfield of artificial intelligence (AI) in which algorithms are used to analyze and learn from data. ML algorithms can improve their performance over time by continuously analyzing new data and adjusting their models accordingly.

Deep learning (DL) is a branch of artificial intelligence that makes use of artificial neural networks for data analysis and interpretation. Based on its capacity to analyze large datasets of dental images and improve the accuracy of diagnosis and treatment planning, deep learning has grown in popularity in dentistry in the past few years. DL has been used in orthodontics to automate the segmentation of dental structures from CBCT scans, which can help with treatment planning and monitoring. In orthodontics, DL has also been used to predict tooth movement and treatment outcomes.

Cone-beam computed tomography (CBCT) is a type of imaging that has revolutionized the field of dentistry by enabling precise diagnosis and treatment planning. With a low radiation dose and a short scanning time, CBCT produces high-resolution three-dimensional images of the dental and maxillofacial structures. CBCT has been widely used in orthodontics, periodontics, endodontics, stomatology, dental implant surgery, maxillofacial surgery, lesion detection, and forensic odontology [1]. However, interpreting these images can be difficult and time-consuming, particularly in complex cases. DL algorithms have been used successfully in various medical fields, including radiology, pathology, and ophthalmology. DL techniques in dentistry have shown promising results in terms of improving CBCT image quality, automating diagnosis, and improving treatment planning.

To explore the most recent developments in Cone Beam Computed Tomography (CBCT) employing AI approaches, a thorough review of research publications released between 2018 and 2022 is carried out. The objective of this paper is to deliver a comprehensive overview of the most recent trends and practices in the use of AI techniques for CBCT in dentistry.

## II. MATERIALS AND METHODS

### A. *Datasets Available for CBCT*

CBCT, or Cone Beam Computed Tomography, is a 3D imaging modality. that can provide detailed information about the teeth, jaw, and facial structures [2]. Collaborating with specialists at various hospitals to collect data while keeping ethical considerations in mind is one of the most effective ways to gather information. Dental CBCT datasets are available from several public sources that can be used in scientific studies. Some examples are as follows:

*1) CTooth:* There are a total of 5803 CBCT slices in the collected data set, and 4243 of those slices have tooth annotations [3]. The images have significant structural differences in terms of tooth position, number, restorations, implants, appliances, and jaw size.

*2) CTooth+:* The CTooth+ 3D dental CBCT dataset was used in the study, which included 146 raw scans and 22 annotated volumes [4]. For future AI-based dental imaging, this study can be used as a benchmark for the tooth volume segmentation task.

*3) The IAN 3D Dataset:* This study talks about a mandibular Cone Beam Computed Tomography (CBCT) dataset that is available to the public and has 2D and 3D manual annotations from expert clinicians [5].

## B. Preprocessing Techniques

In recent years, many CBCT studies have used AI, ML, and DL methods. Preparing the data in a way that the AI algorithm can use it effectively is known as data pre-processing, and it is a crucial step in the AI process for CBCT data. Some common pre-processing methods for CBCT data include:

*1) Noise Reduction:* To remove high-frequency noise, noise reduction techniques like median filtering, wavelet denoising, and 3D Gaussian filter are used.

*2) Cropping the Region of Interest(ROI):* This refers to the manual, automatic, or semi-automatic cropping of the only portion of an image that is of interest.

*3) Intensity Normalization:* Intensity normalization techniques such as histogram equalization, contrast stretching, and scaling is commonly used to standardize and improve image intensity.

*4) CBCT Image Alignment:* It can be challenging to compare and analyze CBCT images due to possible differences in orientation and position. For image alignment of CBCT data, different registration techniques are used.

*5) Augmentation:* Data augmentation techniques such as flipping, rotating, scaling, and adding noise can be used to increase the training dataset size and improve the reliability of the AI algorithm.

# III. OVERVIEW OF AI-BASED CBCT TASKS

Important tasks in CBCT data analysis using AI include detection, classification, and segmentation, with segmentation being largely used in recent years. Different AI architectures have been proposed for these tasks, in the past few years, to improve accuracy and efficiency.

## A. Segmentation

Segmentation is the process of dividing a CBCT image into regions or segments based on their intensity, texture, or shape. For instance, segmenting a tooth from the surrounding tissues. Segmentation algorithms such as 3D-Region Proposal Network (3D-RPN), Mixed-Scale Dense Convolutional Neural Network (CNN MS-D), 3D CNN, Fully Convolutional Network (FCN), Deep Neural Networks (DNN), layered 3D U-Net [6], and Recurrent Neural Network (RNN) are suitable for this task. Unsupervised machine learning algorithms can also be successfully used for CBCT image segmentation. These algorithms learn segmentation rules from CBCT images and annotated training data.

## B. Detection

Detection is the task of identifying specific objects or features in CBCT data. For instance, a CBCT scan can detect the location of a specific tooth or a pathology. DenseNet [7], Dental YOLO, and Dense U-Net are popular object detection algorithms that can perform this task. These algorithms use CNNs to extract features from CBCT images and detect objects of interest.

## C. Classification

Classification is labeling or categorizing a CBCT image or region of interest. For instance, classifying a tooth as healthy or diseased or a CBCT scan as normal or abnormal. Classification algorithms such as support vector machines (SVMs), CNN, and DNN have been used to accomplish this task. These algorithms learn patterns from labeled training data and CBCT image features.

The algorithm of choice for detection, classification, and segmentation depends on the particular need of the application and the properties of the CBCT data. To make sure that the algorithms deliver accurate and dependable results, it is crucial to carefully assess and optimize them.

# IV. APPLICATIONS OF AI IN CBCT

In dentistry, AI has shown promise in a variety of areas, including diagnosis, treatment planning, and oral health monitoring. Figure 1 displays some AI-based CBCT imaging applications. One of the most important uses of AI in dentistry is radiology, where machine learning algorithms can analyze dental images, such as x-rays and CBCT scans to detect and classify abnormalities or lesions. Table 1 presents a summary of AI applications in CBCT in terms of architecture, dataset size, and results.

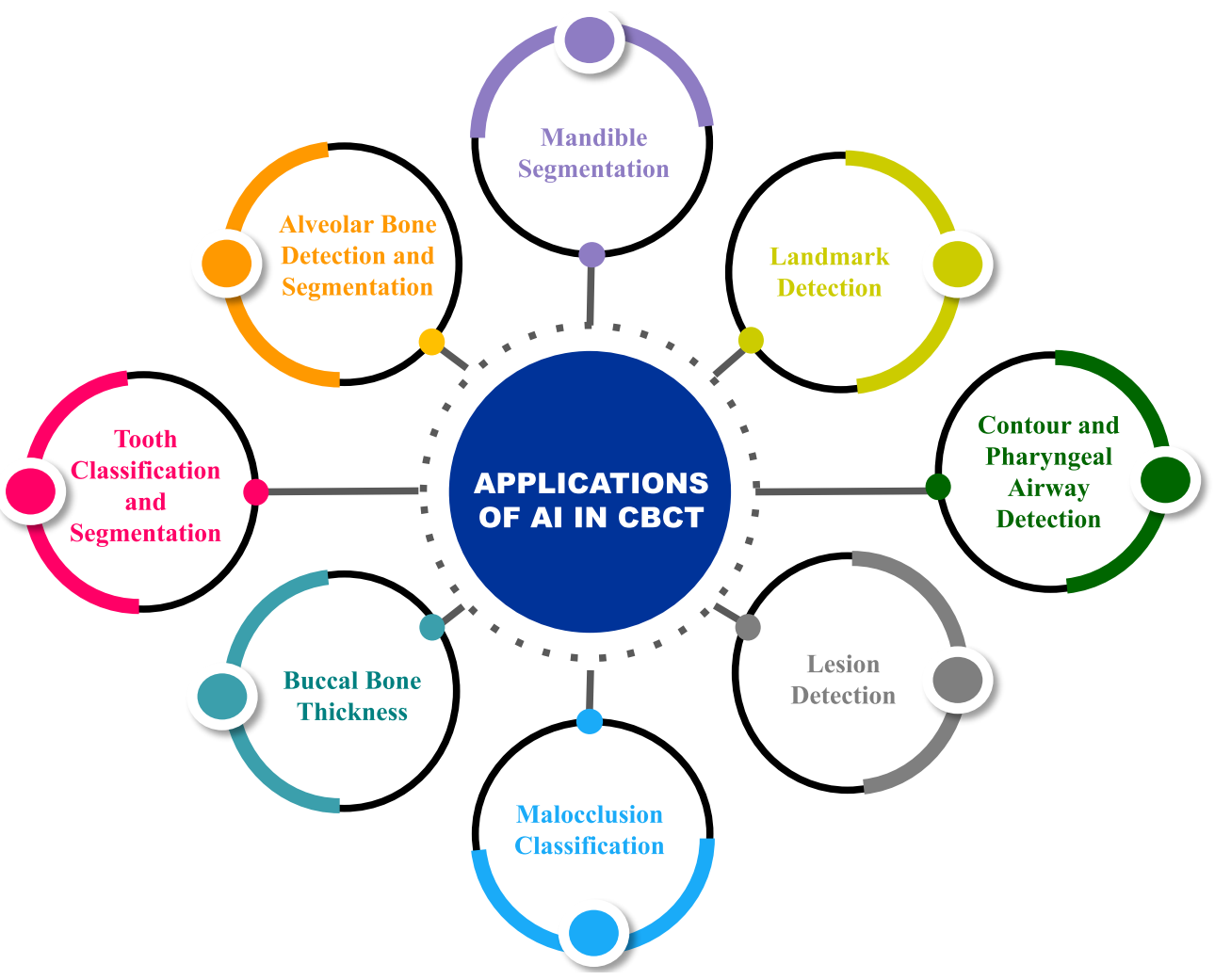

Fig. 1. Applications of AI in CBCT Image Analysis

TABLE I. OVERVIEW OF AI APPLICATIONS IN CBCT (SUMMARY OF ARCHITECTURE, DATASET SIZE, EVALUATION METRIC)

| **Application** | **Architecture** | **Size of Data** | **Evaluation Metric** | **Remarks** |
|---|---|---|---|---|
| Teeth Classification, Segmentation | OCNN | 600 | Classification accuracy = 95.96%, Segmentation accuracy = 89.81% | A novel method to address the issue of misclassification and improved 3-level hierarchical segmentation [8]. |
| Tooth instance identification and segmentation | 3D-RPN | 20 | Detection accuracy = 97.75%, Identification accuracy = 92.79% | A method for automatic and precise instance segmentation for tooth and identification from CBCT images [9]. |
| Anatomical landmarking and segmentation | U-Net and DenseNET, named Tiramisu | 50 | Detection: IoU = 100%, Segmentation: DSC = 93.82% | A deep learning framework for landmark detection and anatomical segmentation [10]. |
| Root morphology assessment | AlexNet, GoogleNet | 760 | Accuracy = 86.9% | A deep learning approach for root morphology assessment to identify whether the distill roots are single or not [11]. |
| Bone segmentation | MS-D network, ResNet, U-Net | 20 | MS-D network: mean DSC = 0.87±0.06, U-Net: mean DSC = 0.87±0.07, ResNet : mean DSC = 0.86±0.05 | A mixed-scale MS-D network for bone segmentation in CBCT scanned images with metal artifacts [12]. |
| Individual tooth segmentation | FCN | 25 | Mean DSC = 0.936 (±0.012) | An approach for individual tooth segmentation that is accurate and automatic for computer-aided analysis [13]. |
| Segmentation of multiclass CBCT | MS-D network | 30 | Jaw: mean DSC = 0.934 ± 0.019, Teeth: mean DSC = 0.945 ± 0.021 | Accurate jaw (mandible and maxilla) and tooth segmentation in CBCT scans for orthodontic diagnosis and treatment planning [14]. |
| Segmentation of alveolar bone and tooth | DNN | 4938 | Tooth: Dice score = 91.5%, Bone: Dice score = 93.0% | An approach for precisely distinguishing individual teeth and alveolar bone for computer-aided analysis and diagnosis [15]. |
| Mandible segmentation | C2FSeg | 59 | Dice score(%) = 95.31 (±1.11) | Accurate segmentation of the mandible from CBCT scans to build a personalized model for orthodontic treatment planning and maxillofacial surgery [16]. |
| Landmark detection and cephalogram synthesis | LeNet-5, ResNet50 | 491 | Detection rate = 86.7% | Automatic landmark detection and cephalometric analysis in CBCT scans [17]. |
| Lesion detection and automatic segmentation | Dense U-Net | 20 | Precision = 0.9, Recall = 0.8 | A method for CBCT segmentation and lesion detection that combines oral-anatomical knowledge with deep learning [18]. |
| Malocclusion Classification | VGG16, Inception-V3 | 218 | VGG16: Accuracy = 93.33%, Inception-V3: Accuracy = 93.83 | Automated skeletal malocclusion detection and classification from 3D CBCT craniofacial images using multi-channel deep learning (DL) models [19]. |

### *A. Tooth Classification and Segmentation*

Tooth classification refers to assigning each tooth in the CBCT image to its corresponding tooth types, such as incisors, canines, premolars, and molars. Tooth segmentation is the process of separating the teeth from the soft tissue and bone around them in a CBCT image. Segmentation measures tooth dimensions and detects dental anomalies like caries and fractures, while classifying teeth to help identify dental pathologies and plan orthodontic treatments. In the past few years, automatic tooth segmentation using AI algorithms has become the focus [20].

### *B. Alveolar Bone Detection and Segmentation*

Detecting alveolar bone involves locating the regions of the CBCT image that correspond to the alveolar bone. After detecting the alveolar bone, it can be segmented by separating it from the surrounding soft tissue and bone. The Dental-YOLO architecture has been studied for the automatic detection of alveolar bone [21].

### *C. Mandible Segmentation*

In CBCT images, mandible segmentation separates the mandible from soft tissue and bone. AI models have been developed for improved and optimized mandibular segmentation [22].

### *D. Landmark Detection*

Landmark detection involves locating anatomical points and features in CBCT images. Orthodontics, implant placement, and maxillofacial surgery require landmark detection. The joint bone Segmentation and landmark Digitization (JSD) framework was put forward to address the relationship between bone segmentation and landmark digitization [23].

### *E. Contour and Pharyngeal Airway Detection*

Contour detection in CBCT images involves identifying the maxilla, mandible, and pharyngeal airway borders. A recent study used possible edge direction for automatic contour detection [24]. The pharyngeal airway at the back of the throat is essential for sleeping and breathing. The pharyngeal airway was automatically segmented from CBCT images using a deep learning algorithm [25].

### *F. Lesion Detection*

Lesions are unusual growths or areas that can signal a disease. Accurate lesion detection in CBCT images has been a focus of research and is still being pursued to facilitate early disease detection and prompt treatment.

### *G. Malocclusion Classification*

A misalignment or incorrect relationship between the teeth of the upper and lower dental arches is referred to as malocclusion. Image processing and machine learning methods have been developed to accurately classify malocclusion.

### *H. Buccal Bone Thickness*

The buccal bone is the outer plate of cortical bone that covers the root of the teeth in the jawbone. Accurate buccal bone thickness measurement in CBCT images can help choose implant sizes and locations and assess implant failure risk due to inadequate bone thickness. Using descriptive and analytical techniques, the impact of menopause on female buccal bone thickness was investigated [26].

## V. BIBLIOMETRIC ANALYSIS

Bibliometrics is the use of visualization & statistical methods to analyze and evaluate scholarly publications by examining their bibliographic information. Bibliometric networks are created and visualized using the software application VOSviewer. These networks can be created using citation, bibliographic coupling, co-citation, or co-authors relationships and can include scholarly articles, researchers, or independent publications. VOSviewer's key feature is its ability to map co-citation and co-occurrence networks. VOSviewer lets users customize visualizations by changing node size, number, color, and more. The visualizations help researchers identify the most important publications, authors, and topics in their field.

### A. *Bibliographic Database Files*

VOSviewer supports five distinct types of bibliographic database files; for this study, the Scopus database was utilized. Search terms like "Deep Learning CBCT Image Analysis" and "Deep Learning CBCT Dental Image" were used to find scholarly articles. The search yielded 125 documents, including 103 research articles, 12 conference papers, 5 review papers, 3 conference reviews, and 2 book chapters. The resulting research documents were exported to a CSV file with pertinent information for visualization such as document title, year, source title, citation count, author, source, and document type, affiliations, author keywords, and original document language.

### B. *Map Visualization*

A map can be created based on network data, which is passed as input in the form of a bibliographic database file (.csv file). VOSviewer allows us to build networks of scientific papers, journals, researchers, research organizations, countries, keywords, or terms. This work includes three visualizations, which are as follows:

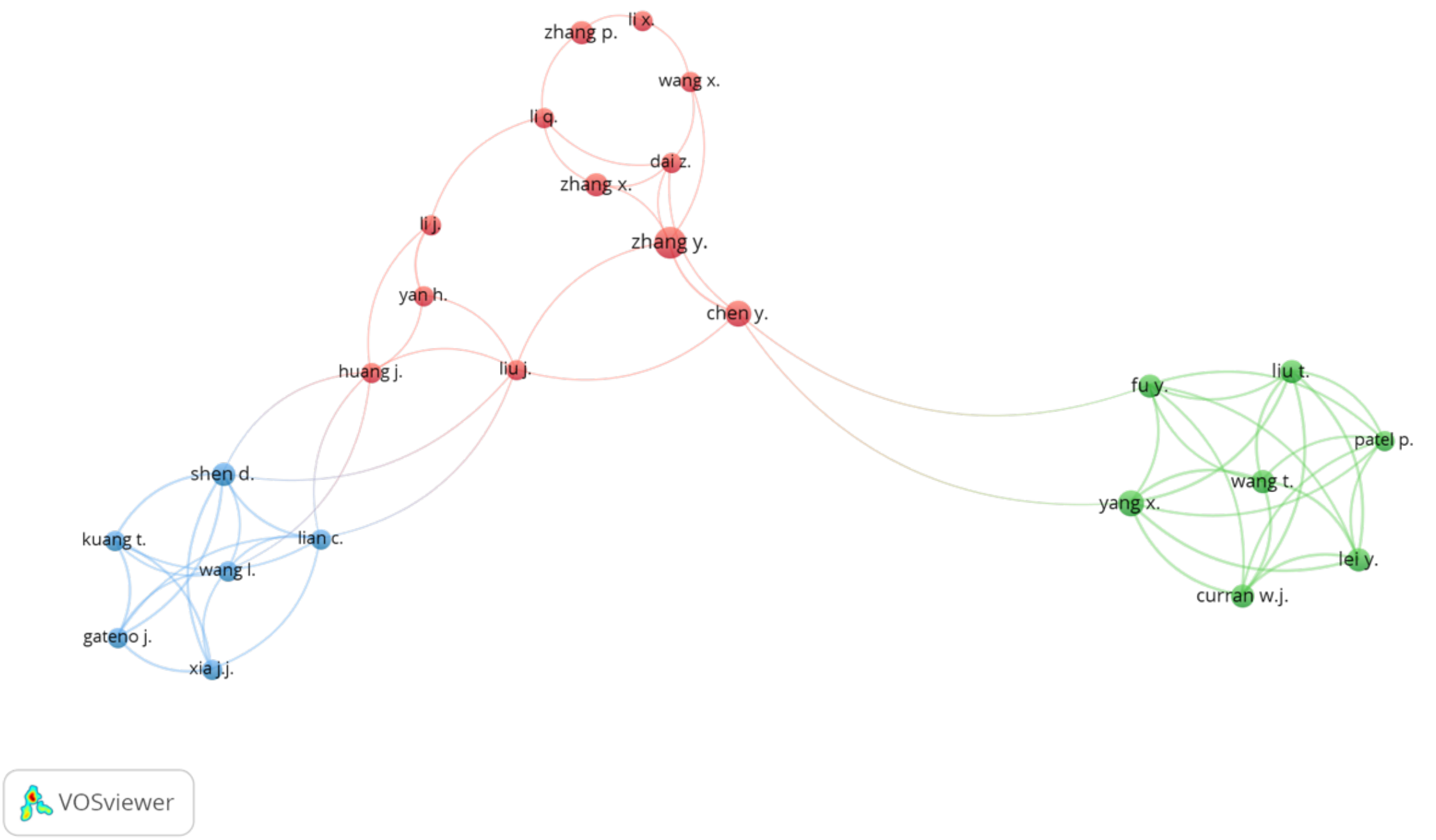

Fig. 2. Map visualization for author and co-author connections. 3 Clusters formed: (a) Cluster 1: Red, Zhang Y. has the most connections (documents: 7, links: 5, link strength: 6), (b) Cluster 2: Green, Yang X. has the most connections (documents: 5, links: 7, link strength: 23), (c) Cluster 3: Blue, Shen D. has the most connections (documents: 4, links: 7, link strength: 16)

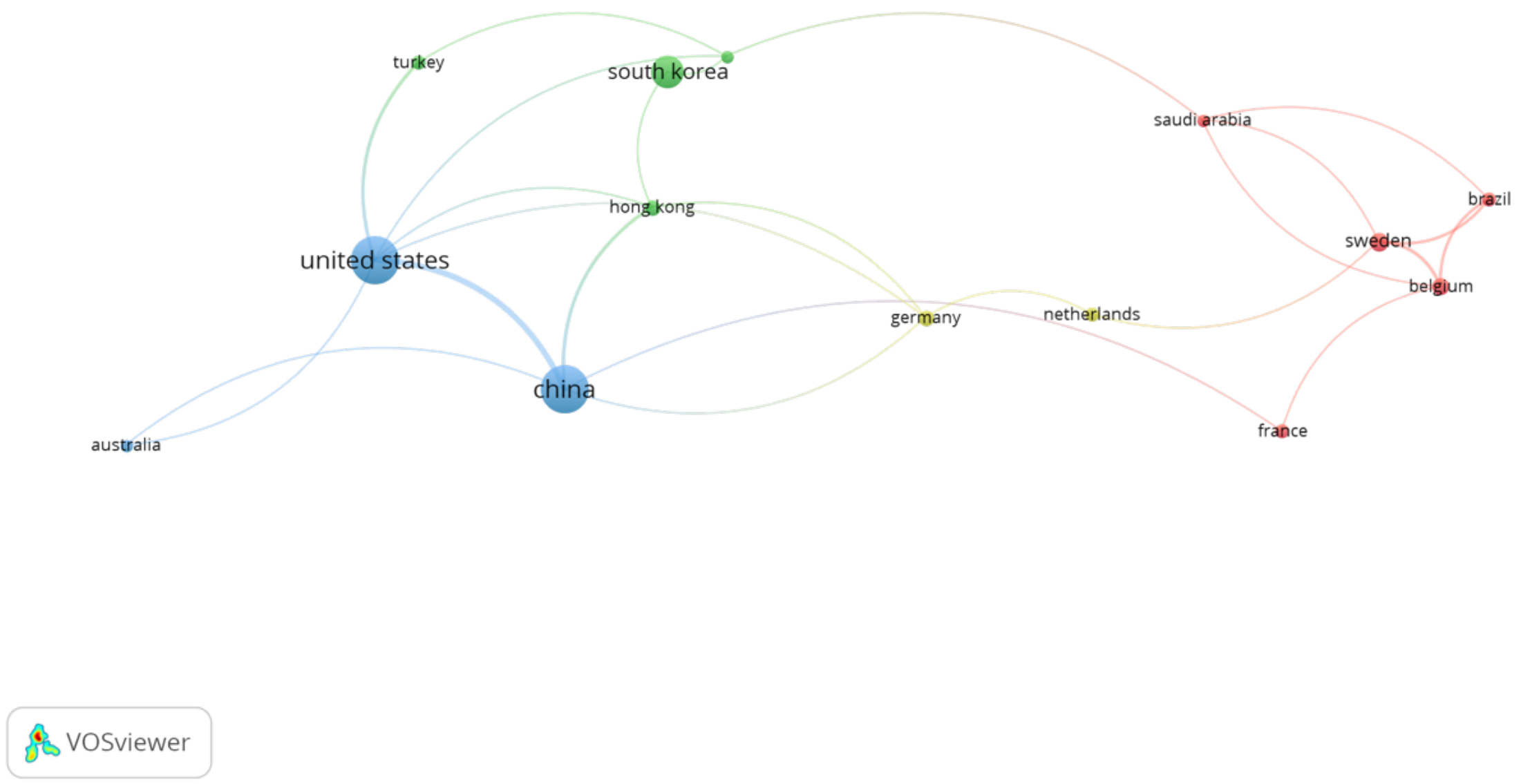

Fig. 3. Map visualization for country and co-author connections. 4 Clusters formed: (a) Cluster 1: Red, Sweden has the maximum number of documents (documents: 7, links: 4, link strength: 9), (b) Cluster 2: Green, South Korea has the maximum number of documents (documents: 19, links: 2, link strength: 2), (c) Cluster 3: Blue, United States has maximum documents (documents: 42, links: 6, link strength: 20), (d) Cluster 4: Yellow, Germany has maximum documents (documents: 5, links: *4, link strength: 4)*

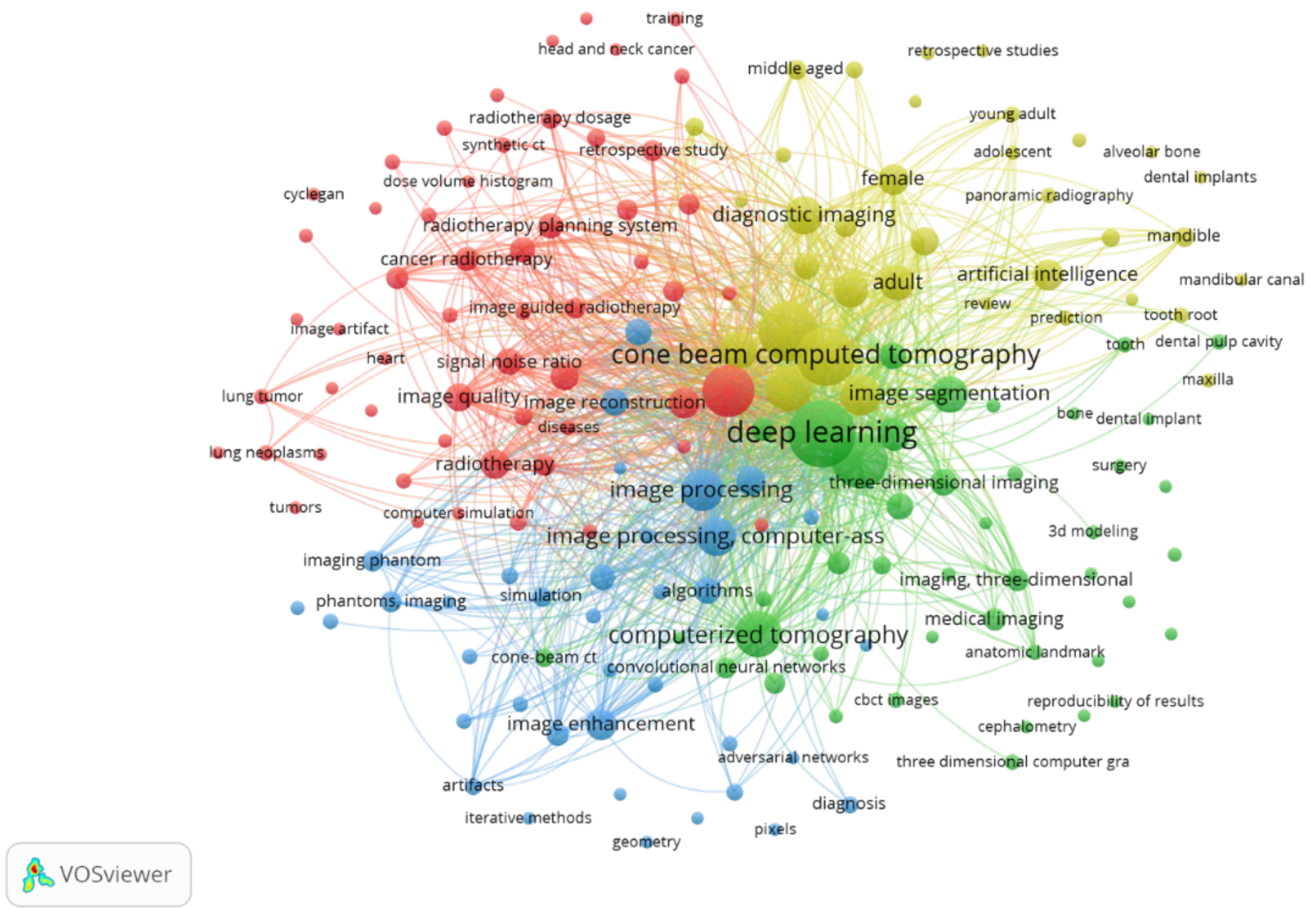

Fig. 4. Map visualization for occurrences of keywords. The most frequent keywords are: (a) deep learning (occurrences: 113, links 171, total link strength: 1968), (b) cone beam computed tomography (occurrences: 88, links 171, total link strength: 1790), (c) computerized tomography (occurrences: 55, links 162, total link strength: 1080), (d) image processing (occurrences: 46, links 165, total link strength: 1044)

*1) Author and Co-author:* The purpose of this visualization was to discover connections between authors and co-authors. To find the connection, various parameters were set: type of analysis = "Co-authorship," unit of analysis = "Author," and counting method = "Full Counting," resulting in a total of 665 authors. For the threshold setting, the minimum number of documents and citations of an author were both set to three, resulting in 41 authors meeting the threshold. Only 25 of the 41 authors were linked, so the map was created. According to the results (Figure 2), the author Zhang Y. has the most connections (documents: 7, links: 5, link strength: 6) of the 25 authors.

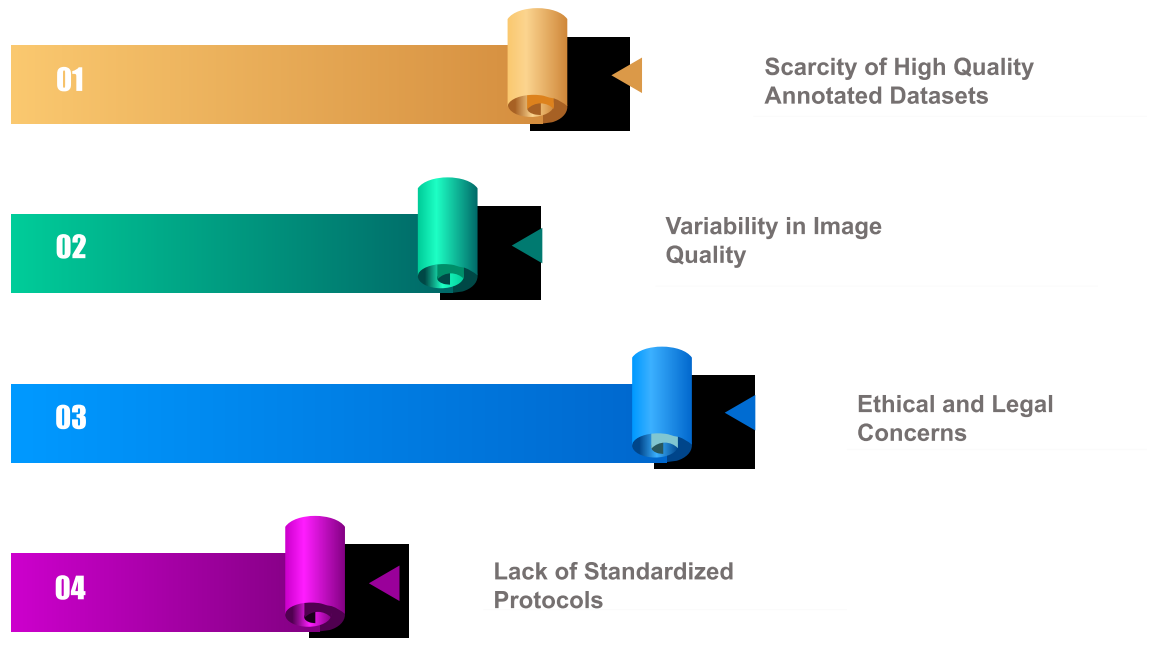

Fig. 5. Challenges and Limitations of AI in CBCT Imaging

*2) Country and Co-author:* The goal of this visualization was to find links between co-authors and countries. To find the connection, the following settings were used: type of analysis = "Co-authorship", the unit of analysis = "Countries", and counting method = "Full Counting", resulting in a total of 41 countries. To determine the threshold, the minimum number of documents and citations for each country was set at 3 and 6, respectively, with the result that 15 countries met the requirement. Only 14 of these countries had connections for which the map was created. According to the results (Figure 3), the United States has the maximum number of documents (documents: 42, links: 6, link strength: 20) of the 15 countries.

*3) Keywords and Co-occurrences:* The objective of this visualization was to find the co-occurrences of all the keywords used in the bibliographic data. The following settings were used to locate the connection: type of analysis = "Co-occurrences", the unit of analysis = "All keywords", and the counting method = "Full Counting", resulting in a total of 1322 keywords. The minimum number of keyword occurrences was set to 4 for threshold setting, resulting in 172 keywords meeting the threshold for which the map was created (Figure 4).

## VI. DISCUSSION AND CONCLUSION

The field of AI has revolutionized the way dental CBCT data is analyzed, and the use of AI techniques in dentistry has significantly increased. Researchers can use publicly available CBCT datasets to test their AI algorithms on real-world data. This has improved the accuracy and efficiency of the algorithms and allowed for the development of new and innovative AI applications in dentistry. Despite the advances in AI for CBCT data, there are still challenges to overcome (Figure 5). One of the most significant challenges is the scarcity of high-quality annotated datasets, which are required for training and validating AI algorithms. Another issue is the lack of standardized protocols for data acquisition and pre-processing, which can cause variability in CBCT data and affect the performance of AI algorithms. In conclusion, AI in CBCT data analysis has the potential to revolutionize dentistry. AI-based algorithms can improve diagnosis and treatment planning by providing accurate and reliable information about dental structures and pathologies. We can expect more innovations in this field in the future. The integration of AI into clinical practice has the potential to improve patient outcomes, reduce costs, and improve overall dental care quality.

## References

[1] H. Gaêta-Araujo *et al.*, "cone beam computed tomography in dentomaxillofacial radiology: A two-decade overview," *Dentomaxillofacial Radiology*, vol. 49, no. 8. 2020. doi: 10.1259/DMFR.20200145.

[2] N. K. Singh and K. Raza, "Progress in deep learning-based dental and maxillofacial image analysis: A systematic review," *Expert Systems with Applications*, vol. 199. 2022. doi: 10.1016/j.eswa.2022.116968.

[3] W. Cui *et al.*, "CTooth: A Fully Annotated 3D Dataset and Benchmark for Tooth Volume Segmentation on Cone Beam Computed Tomography Images," Jun. 2022, [Online]. Available: http://arxiv.org/abs/2206.08778

[4] W. Cui *et al.*, "CTooth+: A Large-scale Dental Cone Beam Computed Tomography Dataset and Benchmark for Tooth Volume Segmentation," Aug. 2022, [Online]. Available: http://arxiv.org/abs/2208.01643

[5] M. Cipriano *et al.*, "Deep Segmentation of the Mandibular Canal: A New 3D Annotated Dataset of CBCT Volumes," *IEEE Access*, vol. 10, pp. 11500–11510, 2022, doi: 10.1109/ACCESS.2022.3144840.

[6] Q. Li, K. Chen, L. Han, Y. Zhuang, J. Li, and J. Lin, "Automatic tooth roots segmentation of cone beam computed tomography image sequences using U-net and RNN," *J Xray Sci Technol*, vol. 28, no. 5, 2020, doi: 10.3233/XST-200678.

[7] Z. Huang, T. Xia, J. Kim, L. Zhang, and B. Li, "Combining CNN with Pathological Information for the Detection of Transmissive Lesions of Jawbones from CBCT Images," in *Proceedings of the Annual International Conference of the IEEE Engineering in Medicine and Biology Society, EMBS*, 2021. doi: 10.1109/EMBC46164.2021.9630692.

[8] S. Tian, N. Dai, B. Zhang, F. Yuan, Q. Yu, and X. Cheng, "Automatic classification and segmentation of teeth on 3D dental model using hierarchical deep learning networks," *IEEE Access*, vol. 7, 2019, doi: 10.1109/ACCESS.2019.2924262.

[9] Z. Cui, C. Li, and W. Wang, "Toothnet: Automatic tooth instance segmentation and identification from cone beam ct images," in *Proceedings of the IEEE Computer Society Conference on Computer Vision and Pattern Recognition*, 2019, vol. 2019-June. doi: 10.1109/CVPR.2019.00653.

[10] N. Torosdagli, D. K. Liberton, P. Verma, M. Sincan, J. S. Lee, and U. Bagci, "Deep Geodesic Learning for Segmentation and Anatomical Landmarking," *IEEE Trans Med Imaging*, vol. 38, no. 4, 2019, doi: 10.1109/TMI.2018.2875814.

[11] T. Hiraiwa *et al.*, "A deep-learning artificial intelligence system for assessment of root morphology of the mandibular first molar on panoramic radiography," *Dentomaxillofacial Radiology*, vol. 48, no. 3, 2019, doi: 10.1259/dmfr.20180218.

[12] J. Minnema *et al.*, "Segmentation of dental cone-beam CT scans affected by metal artifacts using a mixed-scale dense convolutional neural network," *Med Phys*, vol. 46, no. 11, 2019, doi: 10.1002/mp.13793.

[13] Y. Chen *et al.*, "Automatic Segmentation of Individual Tooth in Dental CBCT Images from Tooth Surface Map by a Multi-Task FCN," *IEEE Access*, vol. 8, 2020, doi: 10.1109/ACCESS.2020.2991799.

[14] H. Wang, J. Minnema, K. J. Batenburg, T. Forouzanfar, F. J. Hu, and G. Wu, "Multiclass CBCT Image Segmentation for Orthodontics with Deep Learning," *J Dent Res*, vol. 100, no. 9, 2021, doi: 10.1177/00220345211005338.

[15] Z. Cui *et al.*, "A fully automatic AI system for tooth and alveolar bone segmentation from cone-beam CT images," *Nat Commun*, vol. 13, no. 1, Dec. 2022, doi: 10.1038/s41467-022-29637-2.

[16] B. Qiu *et al.*, "Mandible segmentation of dental cbct scans affected by metal artifacts using coarse-to-fine learning model," *J Pers Med*, vol. 11, no. 6, 2021, doi: 10.3390/jpm11060560.

[17] Y. Huang, F. Fan, C. Syben, P. Roser, L. Mill, and A. Maier, "Cephalogram synthesis and landmark detection in dental cone-beam CT systems," *Med Image Anal*, vol. 70, 2021, doi: 10.1016/j.media.2021.102028.

[18] Z. Zheng, H. Yan, F. C. Setzer, K. J. Shi, M. Mupparapu, and J. Li, "Anatomically Constrained Deep Learning for Automating Dental CBCT Segmentation and Lesion Detection," *IEEE Transactions on Automation Science and Engineering*, vol. 18, no. 2, 2021, doi: 10.1109/TASE.2020.3025871.

[19] I. Kim *et al.*, "Malocclusion Classification on 3D Cone-Beam CT Craniofacial Images Using Multi-Channel Deep Learning Models," in *Proceedings of the Annual International Conference of the IEEE Engineering in Medicine and Biology Society, EMBS*, 2020, vol. 2020-July. doi: 10.1109/EMBC44109.2020.9176672.

[20] P. Lahoud *et al.*, "Artificial Intelligence for Fast and Accurate 3-Dimensional Tooth Segmentation on Cone-beam Computed Tomography," *J Endod*, vol. 47, no. 5, 2021, doi: 10.1016/j.joen.2020.12.020.

[21] M. Widiasri *et al.*, "Dental-YOLO: Alveolar Bone and Mandibular Canal Detection on Cone Beam Computed Tomography Images for Dental Implant Planning," *IEEE Access*, vol. 10, pp. 101483–101494, 2022, doi: 10.1109/ACCESS.2022.3208350.

[22] P. J. Verhelst *et al.*, "Layered deep learning for automatic mandibular segmentation in cone-beam computed tomography," *J Dent*, vol. 114, 2021, doi: 10.1016/j.jdent.2021.103786.

[23] J. Zhang *et al.*, "Context-guided fully convolutional networks for joint craniomaxillofacial bone segmentation and landmark digitization," *Med Image Anal*, vol. 60, 2020, doi: 10.1016/j.media.2019.101621.

[24] I. Marin, I. B. PĂvĂloiu, N. Goga, A. VasilĂţeanu, and G. DrĂgoi, "Automatic contour detection from dental CBCT DICOM data," in *2015 E-Health and Bioengineering Conference, EHB 2015*, 2016. doi: 10.1109/EHB.2015.7391424.

[25] Ç. Sin, N. Akkaya, S. Aksoy, K. Orhan, and U. Öz, "A deep learning algorithm proposal to automatic pharyngeal airway detection and segmentation on CBCT images," *Orthod Craniofac Res*, vol. 24, no. S2, 2021, doi: 10.1111/ocr.12480.

[26] N. Naghibi, K. Fatemi, S. H. Hoseini-Zarch, B. Sadeghi, and M. Fasihi Ramandi, "CBCT evaluation of buccal bone thickness in the aesthetic zone of menopausal women: A cross-sectional study," *Clin Exp Dent Res*, vol. 8, no. 5, pp. 1076–1081, Oct. 2022, doi: 10.1002/cre2.623.